\documentstyle{cupconf}
\input psfig.sty
\begin{document}
  \title[Cambridge University Press]{A photometric study of NGC 458}

  \author[V. Ripepi and E. Brocato]{%
  V\ls I\ls N\ls C\ls E\ls N\ls Z\ls O\ns
  R\ls I\ls P\ls E\ls P\ls I$^1$,\ns 
  \and\ns
  E\ls N\ls Z\ls O\ns 
  B\ls R\ls O\ls C\ls A\ls T\ls O$^2$}
  \affiliation{$^1$Dipartimento di Fisica, 
    Universit\`a della Calabria, Arcavacata di Rende, 
87036 Rende (CS), IT\\[\affilskip]
    $^2$ Osservatorio Astronomico di Collurania, 64100 Teramo, IT}
  \maketitle

\begin{abstract}

We present a CCD investigation of the poorly known SMC young Globular 
Cluster NGC 458. The NTT data, presented here, allowed us to study in 
detail the more internal 
regions of the cluster which are less contaminated by the field. 
On the basis of theoretical isochrones, a preliminary evaluation of 
the age is also given.

\end{abstract}

\firstsection 

\section{Introduction}

The observation of stellar clusters is a relevant tool in testing 
the goodness of stellar evolution theories. In particular when   
intermediate-mass stars (about 2-9 \msun) are investigated, one has to observe 
young clusters. Unfortunately such a clusters in our galaxy, in general, contains 
too few stars to allow significant tests to stellar theoretical models, so 
that it is useful to study young populous clusters in near galaxies; 
in particular in SMC one of the richest cluster is NGC 458, which
appears poorly investigated.
In fact there is no direct measurement of his metallicity or reddening, and 
the estimations of the age (see \cite{hodge83}, \cite{elson85} 
and \cite{sto92}) ranges from 50 to 130 \myr. The only previous 
photometric work on this cluster is due to~\cite{arp59} who 
obtained a photographic BV CM diagram either of the cluster core and of the 
surrounding field. 
For this reasons, NGC 458 was observed and a new CMD for this cluster was
derived.
In the following we will present our preliminary results on NGC 458, giving 
 evaluation for metallicity, reddening and age of this cluster.

\section{Observations, data reduction and the resulting CM diagram}

The CM diagram is based on three B and three V frames obtained with ESO 
NTT telescope equipped with SUSI
(La Silla, Chile) in december 1995. These data was reduced using the ROMAFOT 
package (Buonanno et al. 1979, 1983) for crowded field 
photometry and BV instrumental magnitudes for 1056 stars 
within a radius of 65\arcsec~ were measured. \par 
The small size of the SUSI field of view, 
covering only the  NGC458 core,
was compensated by using a
set of observations based on B V frames obtained at the 0.9 CTIO 
Telescope (Cerro Tololo, Chile), and kindly provided by A. R. Walker. 
The relationships for calibration were derived by  using 9 photoelectric 
standards by~\cite{walk87} and~\cite{alv95} located in the field covered by
CTIO measurements. Common stars 
between NTT and CTIO data were then used to calibrate the NTT photometry.\par
\begin{center}
\begin{figure*}
\vspace{5.7cm}
\special{hoffset=50 voffset=170 hscale=30 vscale=30
angle=-90 psfile=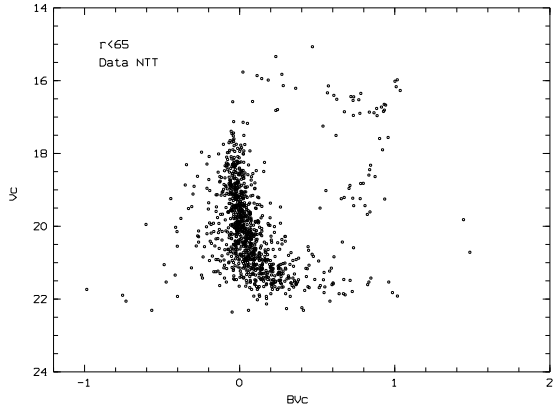}
\caption{NTT-data; CM diagram for the central zone of NGC 458: 
1056 stars in $r<65\arcsec$)}
\label{fig1}
\end{figure*}
\end{center}

 
The NTT-based CM diagram of NGC 458 core ($r<65$\arcsec, 1056 stars) 
is presented in Fig.~\ref{fig1}.  
It shows a well defined MS extending over about four magnitude in \V, 
\V=18.2 \mag~ is the estimated visual magnitude of the TO and a 
consistent population of evolved stars (34 stars) can be recognized 
from  $(\bmv)_0 \simeq -0.1 \mag$ to $(\bmv)_0 \simeq 1 \mag$. \par

\begin{figure*}
\vbox{
\hbox{
\psfig{figure=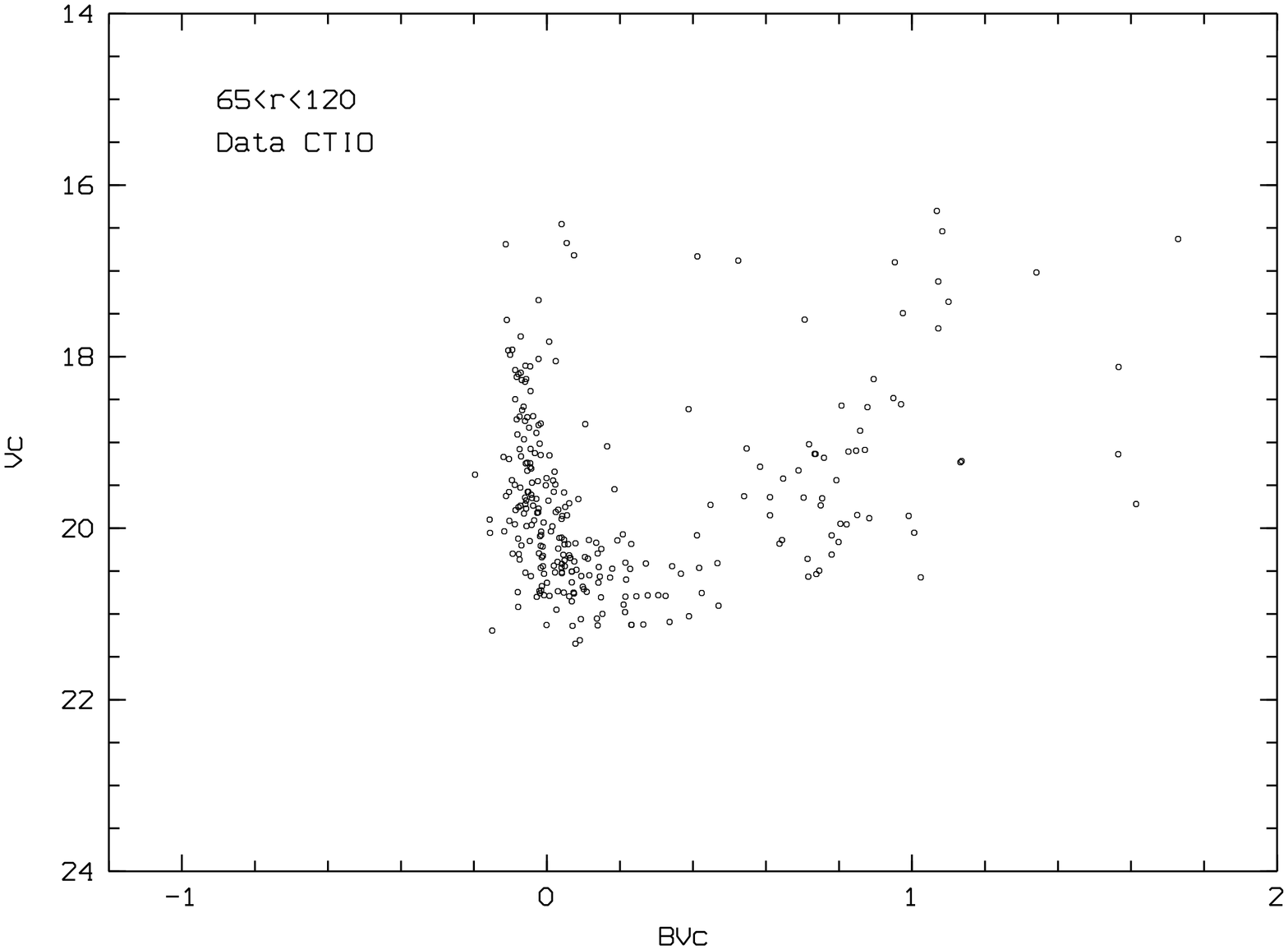,width=7cm,height=6cm,angle=-90}
\psfig{figure=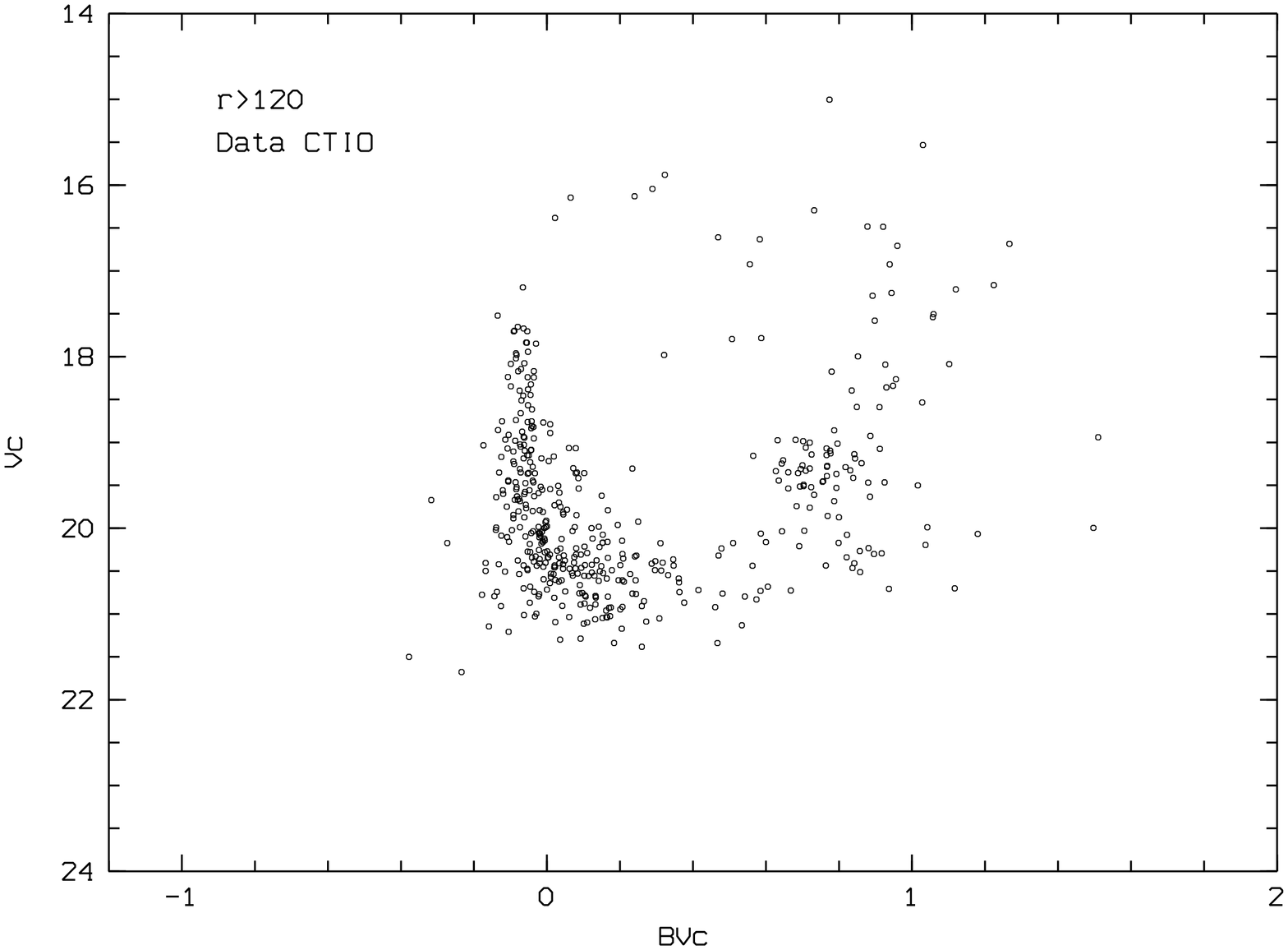,width=7cm,height=6cm,angle=-90}
}}
\caption{CTIO data; left: CM diagram for $65\arcsec<r<120\arcsec$; 
right: $r>120\arcsec$}
\label{fig2ab}
\end{figure*}

The CTIO-based CM diagrams, reported in Fig.s~\ref{fig2ab}, resemble
the main features of the NTT ones. However,  
it can be recognized that, selecting the external region ($r >$ 120\arcsec)
 of the cluster, a clump of field stars appears at 
 $(\bmv)_0 \simeq 1.0 \mag$ and $\V_0 \simeq 19 \mag$. \par
 This clump has been already recognized to be  populated of 
helium burning intermediate mass 
stars belonging to the field (2-3 \msun, see~\cite{cast90}, 
Bencivenni et al. (1991)).\par

It is interesting to note that the field of SMC and LMC appears 
to be very similar according to theoretical previsions discussed in 
~\cite{benc91}.

\section{Results}

Since for NGC 458 there is no indication in literature about reddening we 
first estimated this  quantity. \par
\begin{figure*}[h]
\vbox{
\hbox{
\psfig{figure=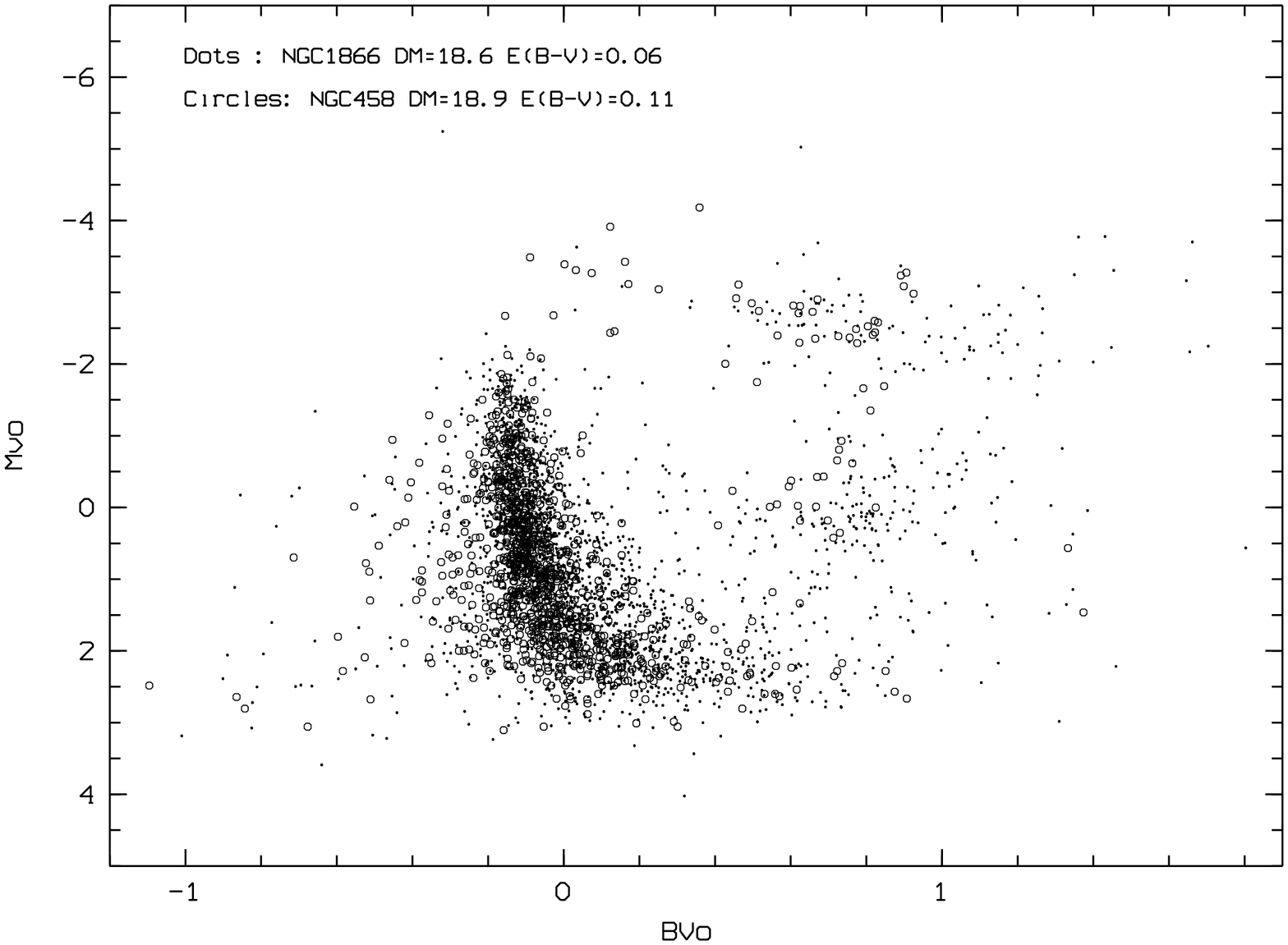,width=7cm,height=7.5cm,angle=-90}
\psfig{figure=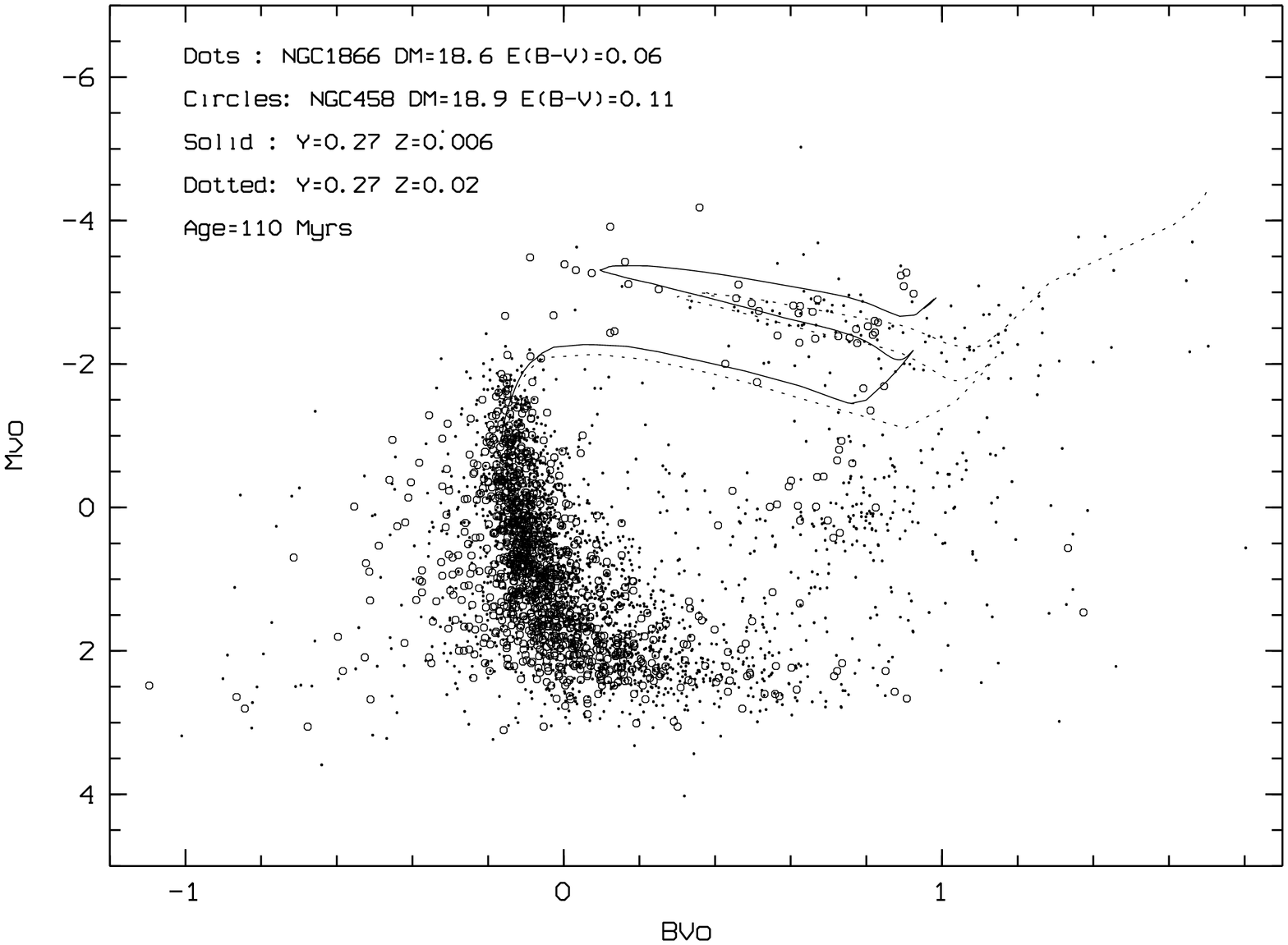,width=7cm,height=7.5cm,angle=-90}
}}
\caption{Comparison between CM diagram of NGC 458 and NGC 1866; Left:
observational data only; Right:as in Left, but with superimposed isochrones  
for 100 \myr but different metallicity}
\label{fig34}
\end{figure*}

To obtain preliminary indication, we compared the CM diagram of NGC 458 with 
that of the  LMC cluster NGC 1866   which has known 
reddening and distance modulus and
presents a quite similar morphology (\cite{broc89}). 
Concerning NGC 458 we assumed the distance 
modulus of SMC, i.e. DM = 18.9 \mag~(\cite{west90}).
Then, by assuming the quoted values of 
distance modulus the two Main Sequences fairly overlap if a reddening 
of E(B-V) = 0.11 is adopted for NGC 458 (Fig.~\ref{fig34} left).
This figure also discloses that
the evolved stars of the two clusters are located at about the same 
luminosity, but NGC 458 evolved stars extend 
to much bluer colors (temperatures). 
Recalling that the temperature extension of the He-burning blue loop 
in intermediate mass  stars depends on metallicity (\cite{broc93}),
a possible explanation is that NGC 458 and NGC 1866 have quite similar ages
(about 110 \myr) but different metallicity (i.e. NGC 1866 
more metal-rich than NGC 458). 
This suggestion is supported by Fig.~\ref{fig34} right, where we overplot 
two isochrones (\cite{cass93}) of different metallicities (Z=0.02 and Z=0.006)
but same ages (110\myr)
on the observational data. \par
\begin{figure*}[t]
\vbox{
\hbox{
\psfig{figure=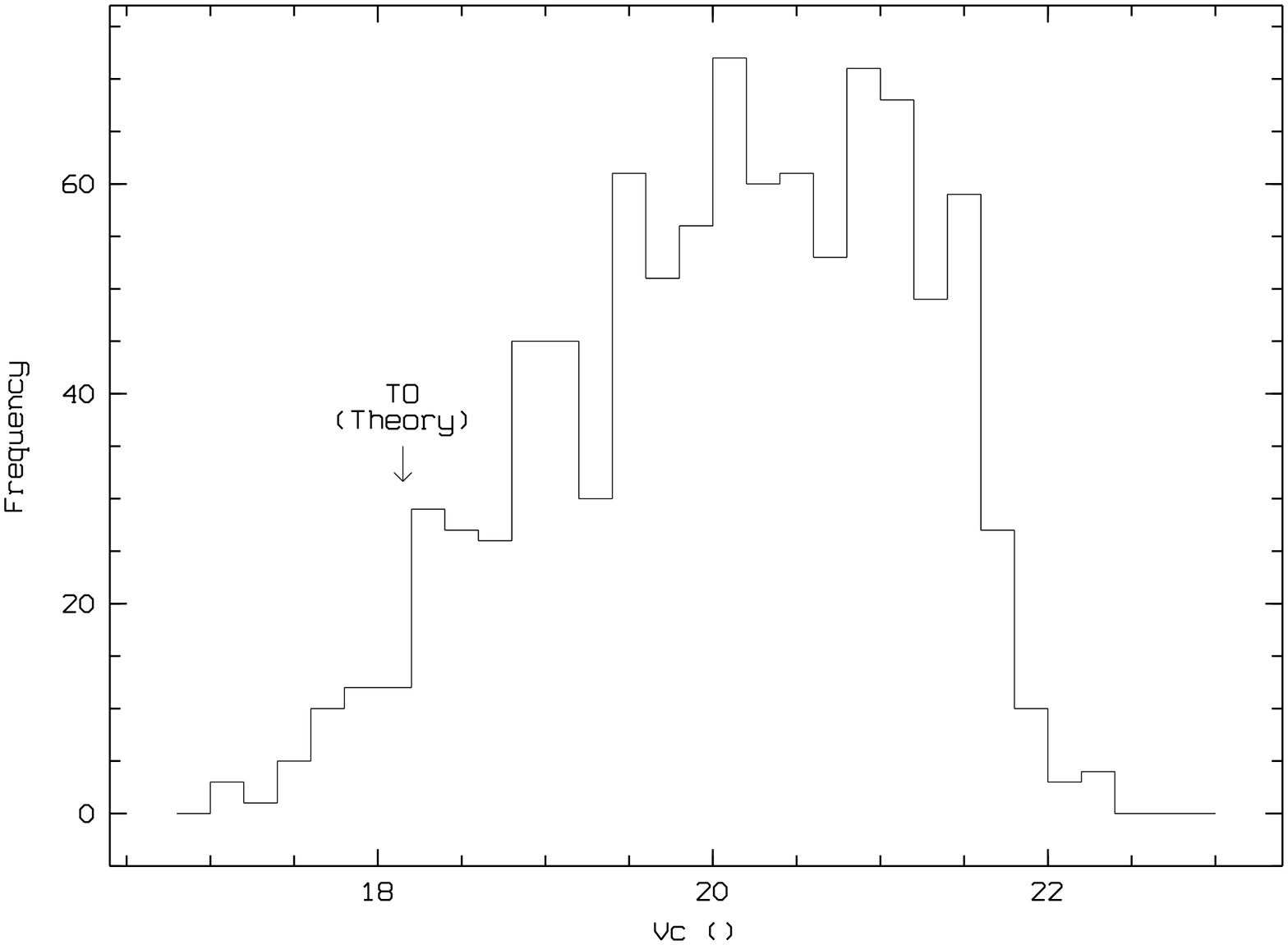,width=7cm,height=6cm,angle=-90}
\psfig{figure=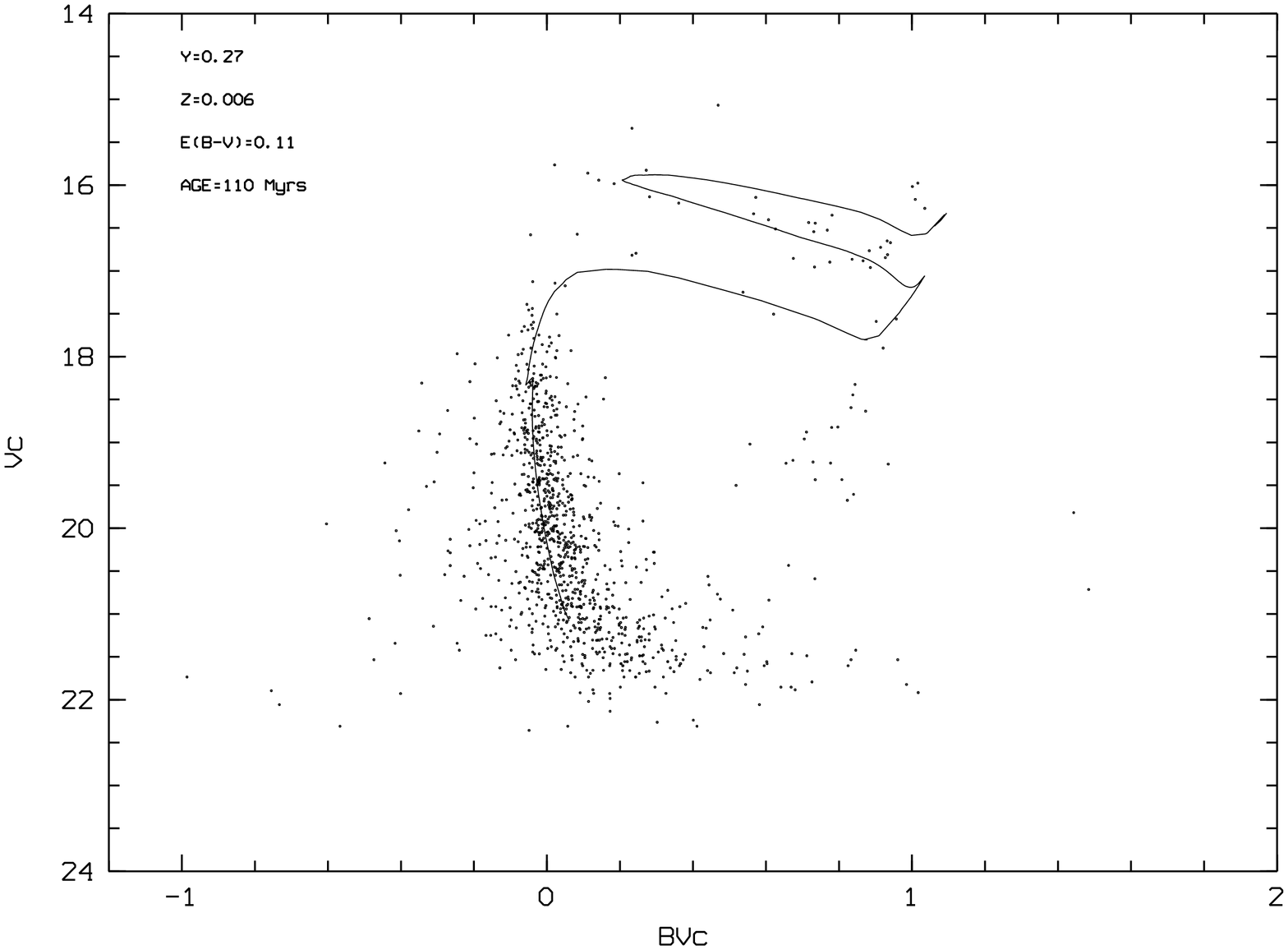,width=7cm,height=6cm,angle=-90}
}}
\caption{Left: Observed LF for NGC 458 with superimposed the 
location of the theoretical TO for an age of 110 \myr; 
Right: the adopted isochrone fitting}
\label{fig56}
\end{figure*}

We also checked the consistency of the age evaluation by 
plotting  (Fig.~\ref{fig56} left) the observed LF of NGC 458 to compare the
observed bright MS termination end to the theoretical one
obtained  by the quoted isochrone of  t = 110 \myr.\par
As a conclusion we find that a fairly good isochrone fitting 
for NGC 458 (Fig.~\ref{fig56} right) is reached with the following values:
DM=18.9, \ebmv=0.11\mag, t=110\myr~ and Z=0.006.

\par 
Independent evaluation of metallicity and reddening would certainly 
improve the reliability of the values derived in this work 
for the first time and which should be regarded as temptative estimates.

\section{Conclusions}

We presented a CM diagram for NGC 458 mainly based on NTT data; we are 
able to fairly identify the various stages of stellar evolution present 
in this cluster; in particular the MS and the blue loop of the evolved stars
are clearly defined.\par
On the basis of our diagram and by comparison with observational data of 
NGC 1866 in LMC we estimated the reddening (\ebmv=0.11\mag) of NGC 458 and
we suggest that these two cluster have very similar age (110 \myr) 
but different metallicity, being NGC 458 less metal-rich. \par
Moreover,  theoretical isochrones provide a fairly
good fit assuming the following  quantities:
Z=0.006, \ebmv=0.11\mag, age=110\myr.\par
As a final remark we have to note that more work is still necessary 
in order to improve and refine the interpretation of  NGC 458: in particular 
independent evaluation of reddening and metallicity would greatly improve
the reliability of the age determination.

\section{Acknowledgments}

We are in debt with A.R. Walker that kindly provided us the CTIO data.

{}

\end{document}